\RequirePackage{fix-cm}
\documentclass[smallcondensed,onecolumn,a4]{svjour3}       % onecolumn (second format)
\smartqed  % flush right qed marks, e.g. at end of proof
\usepackage{graphicx,pdfpages}
\begin{document}

\title{The INPOP10a planetary ephemeris and its applications in fundamental physics.
}
%\subtitle{Do you have a subtitle?\\ If so, write it here}

%\titlerunning{Short form of title}        % if too long for running head

\author{A. Fienga         \and        
		J. Laskar \and
	P. Kuchynka \and
        H. Manche \and
        G. Desvignes \and
 	M. Gastineau \and
	I. Cognard \and
	G. Theureau
}

%\authorrunning{Short form of author list} % if too long for running head

\institute{A. Fienga \at
              Institut UTINAM-CNRS 6213, Besancon, France \\
              Tel.: +0033-381666925\\
%              Fax: +123-45-678910\\
              \email{agnes.fienga@obs-besancon.fr}           %  \\
%             \emph{Present address:} of F. Author  %  if needed  
           \and
          J. Laskar \at
              IMCCE-CNRS 8028, Paris, France\\
		\email{laskar@imcce.fr}  
		           \and
          P. Kuchynka \at
              IMCCE-CNRS 8028, Paris, France\\
		\email{kuchynka@imcce.fr}  
		\and
           H. Manche \at
              IMCCE-CNRS 8028, Paris, France\\
		\email{herve.manche@imcce.fr}  
		\and
	G.Desvignes  \at
              LPCE-CNRS 6115, , France\\
		\email{gdesvignes@astro.berkeley.edu}  
		\and
	M. Gastineau\at
              IMCCE-CNRS 8028, Paris, France\\
		\email{gastineau@imcce.fr}  
		\and
	I. Cognard \at
              LPCE-CNRS 6115, France\\
		\email{icognard@cnrs-orleans.fr}  
		\and
	G. Thereau \at
              LPCE-CNRS 6115, Paris, France\\
		\email{gilles.thereau@cnrs-orleans.fr}  
		\and
}

\date{Received: 19 April 2011 / Accepted: 16 August 2011}
% The correct dates will be entered by the editor

\maketitle

\begin{abstract}
Compared to the previous INPOP versions, the INPOP10a planetary and lunar ephemeris has several improvements. For the planets of our solar system, no big change was brought in the dynamics but improvements were implemented in the fitting process, the data sets used in the fit and in the selection of fitted parameters. 
We report here the main characteristics of the planetary part of INPOP10a like the fit of the product of the Solar mass with the gravitational constant (GM$_{\odot}$) instead of the astronomical unit. 
Determinations of PPN parameters as well as adjustments of the Sun J2 and of asteroid masses are also presented.
New advances of nodes and perihelia of planets were also estimated and are given here.
As for INPOP08, INPOP10a provides to the user, positions and velocities of the planets, the moon, the rotation angles of the  Earth and the Moon as well as  TT-TDB chebychev polynomials at {\it{http://www.imcce.fr/inpop}}.

% \PACS{PACS code1 \and PACS code2 \and more}
% \subclass{MSC code1 \and MSC code2 \and more}
\end{abstract}

\section{Introduction}

Since INPOP08 new observations have been produced, in particular normal points deduced from the three flybys of the Messenger spacecraft around Mercury. These data were the first accurate positions of Mercury since the two Mariner flybys in the seventies. 
Furthermore, the asteroid masses obtained during the fit of INPOP08 (Fienga et al. 2009 \cite{2009A&A...507.1675F}) to observations did not seem to be satisfactory. The method used for these estimations has been revised and a more constraining approach was tested.
The question of fitting the mass of the Sun instead of the astronomical unit in the planetary ephemerides was discussed in the community. We decided to test the possibility of such adjustment by building an ephemerides with a fixed value of AU and a fitted value of GM$_{\odot}$. 
Finally, crucial normal points for the outer planet orbits missed in the INPOP08 data sets have been added. 

In this paper, we introduce the planetary ephemerides INPOP10a and some applications. As no modification was brought to the dynamical modeling, we first describe the main input to the observational data sets used for the adjustment. 
We then introduce the modifications implemented in the selection of used constants and fitted parameters.
In the section 4, we describe the obtained results in term of comparisons to other planetary ephemerides (INPOP08, DE421 (Folkner et al. 2008 \cite{DE421})) and of postfit comparisons to observations. 
The first combined estimations of PPN parameters $\beta$ and $\gamma$ based on INPOP are presented here as well as a list of 58 asteroid masses estimated with INPOP10a.
Supplementary advances of perihelia and nodes of planets are also given.
Finally we introduce the first estimation of rotation angles between planetary ephemerides frames and the ICRF deduced from the VLBI and radio astrometry of millisecond pulsars. 
We check the internal mas-level accuracy of the  Earth orbit but we detect a possible 10 milliarcseconds (mas) rotation between all the planetary solutions and the ICRF. This result needs to be verified after the densification of pulsars with a mas-level VLBI and radio astrometry. 

\section{The INPOP10a Data Sets}
\label{section2}

A detailed description of the observations used for the construction of INPOP10a can be found in (Fienga et al. 2010 \cite{2010journees}).
Several data sets have been added since INPOP08. The global observational distribution for the INPOP fit has changed its balance compared to INPOP06: now, more than 56$\%$ of the planetary observations are deduced from the tracking data of spacecrafts including range, VLBI angular positions and flyby normal points.
The statistics of the obtained postfit residuals as well as the number of points and their distribution in time are presented in table \ref{omctab}.

 The first spacecraft flying by Mercury was Mariner 10. From this mission and its two flybys of Mercury in 1974 and 1975 were extracted two corrections to the geocentric distance of the planet, provided by JPL (Folkner 2010 \cite{F2010}) with an accuracy of about 150 meters.
The MErcury Surface, Space ENvironment, GEochemistry, and Ranging (MESSENGER) spacecraft, only the second probe to encounter the innermost planet, flew by Mercury on 14 January 2008, 6 October 2008, and in December 2009. 
   
The MESSENGER payload includes the Mercury Laser Altimeter                              
(MLA) (Cavanaugh et al. 2007 \cite{2007SSRv..131..451C}) and a gravity science investigation                        
that uses the spacecraft tracking system (Srinivasan et al. 2007 \cite{2007SSRv..131..557S}).                         
During the flybys the spacecraft was tracked at                         
X-band frequency by the Deep Space Network (DSN), the latter                                
providing high-quality Doppler range-rate observations of the                              
spacecraft motion relative to  Earth. 
From the three flybys and the accurate orbit determination obtained by (Smith et al. 2010 \cite{2010Icar..209...88S}), we extracted three corrections to Mercury geocentric positions with an accuracy of about a few mas in angles (right ascension and declination) and a few meter in Earth-Mars range. 
The MESSENGER orbit reconstructed from tracking data by the mission
navigation team is available through the NASA Planetary Data System and can
be accessed with SPICE software (Turner 2007 \cite{Turner2007}). With
the Mercury-centric orbit of the spacecraft, the navigation team provides
also orbits of planets (Earth, Mercury and Venus) fitted only during the flybys of the spacecraft. 
We then deduced for each flyby of Mercury, corrections to Mercury's
geocentric orbit in angular positions and distances. 
%The orbit of Mercury is thus described by each
%correction (also called normal point) with an accuracy of a few milliarcsecond in right ascension and declination and of a few meter in terms of Earth-Mercury range. 

These five points changed our knowledge of the Mercury orbit: until now, only direct radar ranging on Mercury's surface were available with an accuracy of about 800 meters. 

For Mars and Venus, like with INPOP08,  tracking data of MEX and VEX missions  provided by ESA (Morley 2009 \cite{M2009}, Morley 2010 \cite{M2010}) are used in the fit as described in Fienga et al. (2009) \cite{2009A&A...507.1675F}. 
In addition to the Saturn Cassini normal points provided by JPL over the 2005 to 2007 period and used in INPOP08, are added VLBI observations of the spacecraft (Jones et al. 2011 \cite{2011AJ....141...29J}) with an accuracy better than a few mas. 
These VLBI differential positions of Cassini related to ICRF sources were obtained during its mission about Saturn and its satellites.

Flybys data of Jupiter, Uranus and Neptune obtained during several missions (Pioneer 10 and 11, Viking 1 and 2, Ulysses and Cassini) were also added, provided by Folkner (2009) \cite{F2009}. 
These observations improve the estimates of the geocentric distances of the outer planets while no observations of similar type were used in INPOP06 and INPOP08 adjustments.
New optical data obtained from 2000 to 2008 with the Flagstaff Astrometric Scanning Transit Telescope are also added for Uranus, Neptune and Pluto (Folkner 2009 \cite{F2009}).
Stellar occultations (Sicardy 2009 \cite{S2009}) are taken into account in INPOP10a by the use of measured offsets in topocentric ($\alpha$,$\delta$).

%A detailed description of the fit procedure for the moon orbit and libration and of the Lunar Laser Ranging observations used for the fit is presented by (Manche et al. 2010) in this volume.

\begin{table}
\caption{Statistics of the residuals obtained after the INPOP10a fit. For comparison, means and standard deviations of residuals obtained with INPOP08 are given. }
\begin{center}
\begin{tabular}{l l l | c c | c c }
\hline
\small{Planet} & & & \multicolumn{2}{c}{\small{INPOP08}} & \multicolumn{2}{c}{\small{INPOP10a}}\\
%\small{Type of Data} & Nbr& Time & & & &\\
 \small{Type of Data} & Nbr& Time Interval &  mean &1$\sigma$& mean & 1$\sigma$\\
\hline
%\small{Mercury range [m]} & & & & \\
\small{Mercury} & & & & & &\\
\small{Direct range [m]} & 462 & 1965-2000 & 30 & 842 &  7 & 866\\
\small{Mariner range [m]}& 2 & 1974-1975& -1000 & 305 & -28 & 85\\
\small{Messenger Mercury flybys} & & & & \\
\small{Messenger ra [mas]} & 3 & 2008-2009 &  1.1 & 0.7  & 0.4 & 1.2\\
\small{Messenger de [mas]} & 3 & 2008-2009 & 2.0 & 1.9 & 1.9 & 2.1\\
\small{Messenger range [m]} & 3 & 2008-2009 & 52 & 619 & -0.6 & 1.9\\
\hline
\small{Venus } & & & & & &\\
%[m] & &  \\
\small{ Direct range [km]} & 489 & 1965-2000 &  0.5 & 2.3 &  0.5 & 2.2\\
\small{ VEX range [m]} & 22145 & 2006-2010 & 1.6 & 4.4 &  -0.2 & 3.9\\
\small{ VLBI [mas]} & 22 & 1990-2007 & 2 & 2 &   2 & 2.5 \\
\hline
\small{Mars range [m] } & & & & & &\\
%[m] & &  \\
\small{MGS} & 10474 & 1998-2008 & -0.9 & 1.6 &   0.5 & 1.9 \\
\small{MEX} & 24262 & 2006-2010 & -3.5 & 2.0 &    0.0 & 1.7 \\
\small{Path} & 90 & 1997 & 6.8 & 12.5 & -5.0 & 5.0 \\
\small{Vkg} & 1256 & 1976-1982 & -27.4 & 19.0 & -5.7 & 35.0 \\
\small{Mars VLBI [mas]} & 96 & 1989-2007 & 0.5 & 0.5 & -0.0 & 0.4\\
\hline
\small{Jupiter flybys} & & & & & &\\
ra [mas] & 5 & 1974-2000 &48.0 & 40.0  &  6 & 5 \\
de [mas] & 5 &  1974-2000 &-10.0 & 50 &  -13 & 18 \\
\small{range [km]} & 5 & 1974-2000 &  -27 & 55 &  -0.6 & 1.6 \\
\small{Jupiter VLBI} [mas] & 24 & 1996-1997&  4 & 11 & 0.2 & 11\\
\small{Jupiter Optical} & & & & \\
ra [mas] & 6216 & 1914-2008 &  20 & 304 & -26 & 304\\
de [mas] &  6082 & 1914-2008 &  -44 & 313 & -54 & 303\\
\hline
\small{Saturn Cassini} & & & & & &\\
ra [mas] & 31 & 2004-2007 & 1.5 & 4  &  0.7 & 4 \\
de [mas] &  31 & 2004-2007 & 7.0 & 7 &  6.5 & 7 \\
\small{range [m]} & 31 & 2004-2007 &  0.5 & 22 &  0.0 & 17 \\
\small{Saturn VLBI Cassini} & & & & \\
ra [mas] & 10 & 2004-2009 & 0.3 & 0.7 &  0.0 & 0.6 \\
de [mas] & 10 & 2004-2009 &  -1.2 & 2.0 &  0.1 & 0.4 \\
\small{Saturn Optical} & & & & & & \\
ra [mas] & 7824 & 1914-2008 &  -16 & 305 & -16 & 305  \\
de [mas] & 7799 & 1914-2008 &  -7 & 276 & -9 & 276   \\
\hline
\small{Uranus flybys} &  & & & & &\\
ra [mas] & 1 & 1986 &-90&  &  -30 &\\
de [mas] & 1 &  1986 &-36& &  -7 &\\
\small{range [km]} & 1 &  &  1139 & & 0.080& \\
\small{Uranus Optical} & & & & \\
ra [mas] & 4145 & 1914-2008 &   -44 & 278  &  -27 & 290  \\
de [mas] & 4130  & 1914-2008 &   -38 & 339  &  -11 & 338   \\
\hline
\small{Neptune flybys} & & & & & &\\
ra [mas] & 1 & 1989 &-88  &  &-11 &\\
de [mas] & 1 &  1989 &-48 &  &-10 &\\
\small{range [km]} & 1 &  &  2305 & &  0.004 &\\
\small{Neptune Optical} & & & & \\
ra [mas] & 4340 & 1914-2008 &  -32 & 282 & 2 & 281  \\
de [mas] & 4320 & 1914-2008 &  -36 & 335 & 2 & 330   \\
\hline
\small{Pluto occultation} & & && & &\\
ra [mas] & 13 & 2005-2009 &  -6 & 46 & -1 & 47  \\
de [mas] & 13 & 2005-2009 &  16 & 30 & -2 & 19   \\
\small{Pluto Optical} & & && & &\\
ra [mas] & 2449 & 1914-2008 &  353 & 926 & 38 & 629  \\
de [mas] & 2463 & 1914-2008 &  -22 & 524 & 17 & 536   \\
\hline
\end{tabular}
\label{omctab}
\end{center}
\end{table}

\begin{table}
\caption{Asteroid masses found in the recent literature and compared to the values estimated in INPOP10a. [1]: (Baer et al 2011 \cite{2011Icar..212..438B}), [2]: (Baer et al 2008 \cite{2008CeMDA.100...27B}), [3]: (Marchis et al. 2008a \cite{2008Icar..196...97M}), [4]: (Marchis et al. 2008b \cite{2008Icar..195..295M}), [5]:(Yeomans et al. 1997 \cite{1997Sci...278.2106Y}), [6]: (Kochetova 2004 \cite{2004SoSyR..38...66K}). K11 stands for (Konopliv et al. 2011) \cite{2011Icar..211..401K}. [*]: INPOP08e. The last column gives the impact of each asteroid on the  Earth-Mars distances over the 1970 to 2010 period. The impact is defined as the difference in meters between two integrations of the Mars geocentric orbit, with and without the corresponding asteroid. The uncertainties for INPOP10a and INPOP08 are given at 1 formal sigma deduced from least squares. For the close-encounters and K11 determinations, the uncertainties are given at 1 published sigma}
\begin{center}
\begin{tabular}{l l c l l l l}
\hline
\small{Asteroids} & \small{INPOP10a} & \small{Others} & & \small{INPOP08} & \small{K11} & \small{Impact }\\
& & & & & &1970-2010\\
IAU& $10^{12} \times M_{\odot}$ & $10^{12} \times M_{\odot}$ && $10^{12} \times M_{\odot}$ & $10^{12} \times M_{\odot}$ & m\\
\hline
1 & 475.8 $ \pm $ 2.8 & 475.700 $ \pm $ 0.72 &[1] &465.8 $ \pm $ 0.9 & 467.900 $ \pm $ 3.250 & 794\\
2 & 111.4 $ \pm $ 2.8 & 101.000 $ \pm $ 6.5 & [1]&107.6 $ \pm $ 2.0 & 103.440 $ \pm $ 2.550 & 146\\
4 & 133.1 $ \pm $ 1.7 & 130.00 $ \pm $ 0.53 & [1]&132.9 $ \pm $ 3.0 & 130.970 $ \pm $ 2.060 & 1199\\
7 & 7.7 $ \pm $ 1.1 & 8.12 $ \pm $ 0.46 & [1] &5.0 $ \pm $ 0.2 & 5.530 $ \pm $ 1.320 & 28\\
324 & 4.67 $ \pm $ 0.38 & & & 5.6 $ \pm $ 0.08 & 5.340 $ \pm $ 0.990 & 94\\
3 & 11.6 $ \pm $ 1.3 & 14.400 $ \pm $ 2.3 & [1] & 7.5 $\pm$ 0.3 & 12.100 $ \pm $ 0.910 & 56\\
6 & 7.1 $ \pm $ 1.2 & 6.40 $ \pm $ 0.67 &[1] & 0.16 $ \pm $ 0.11 & 6.730 $ \pm $ 1.640 & 21\\
8 & 4.07 $ \pm $ 0.63 & 3.33 $ \pm $ 0.42 & [1] & 5.35 $ \pm $ 0.5 & 2.010 $ \pm $ 0.420 & 13\\
9 & 5.700 [1] & 5.700 $ \pm $ 1.1 & [1] & 1.15 $ \pm $ 1.0 & 3.280 $ \pm $ 1.080 & 30\\
10 & 44.500 [2] & 43.58 $ \pm $ 0.74 & [1] &  & 44.970 $ \pm $ 7.760 & 77\\ 
11 & 1.9 $ \pm $ 1.0 & 3.090 $ \pm $ 0.989 &[2] &  &  &17\\
13 & 8.200 [2] & 8.00 $ \pm $ 2.2 & [1] &  &  & $<$1\\
14 & 4.130 [2] & 3.49 $ \pm $ 0.82 & [1] & 0.71 $\pm$ 0.1  & 1.910 $ \pm $ 0.810 & 18\\
15 & 18.8 $ \pm $ 1.6 & 15.597 $ \pm $ 0.15 & [1] & 22.30 $ \pm $ 1.80 & 14.180 $ \pm $ 1.490 & 22\\
16 & 11.2 $ \pm $ 5.2 & 11.40 $ \pm $ 0.42 & [1] & 15.96 $\pm$ 0.32 & 12.410 $ \pm $ 3.440 & 10\\
18 & 1.845 [2] &  & &  &  & 11\\
19 & 6.380 [2] & 4.18 $ \pm $ 0.36 & [1]& 2.02 $ \pm $ 0.20 & 3.200 $ \pm $ 0.530 & 59\\
20 & 2.850 [2] & 1.68 $ \pm $ 0.35 & [1]& &  & $<$ 1\\
21 & 1.3 $ \pm $ 1.2 & 1.31 $\pm$ 0.44 & [1] & 1.034 $ \pm $ 0.30 & & 5 \\
24 & 2.8 $ \pm $ 1.9 & 5.670 $ \pm $ 2.155 & [2] & &  & 26\\
25 & 0.002 $ \pm $ 0.002 &  & &0.301 $ \pm $ 0.030 &  & 3\\
28 & 4.65 $ \pm $ 1.0 &  & &&  & 4\\
29 & 5.920 [2] & 7.63 $ \pm $ 0.31 & [1] & 4.91 $ \pm $ 0.90 & 7.420 $ \pm $ 1.490 & 27\\
31 & 3.130 [2] & 2.92 $ \pm $ 0.99 & [1] & 29.9 $ \pm $ 6.8 &  & 23\\
41 & 9.2 $ \pm $ 2.6 &  && 5.27 $ \pm $ 0.5 & 4.240 $ \pm $ 1.770 & 12\\
42 & 1.8 $ \pm $ 1.07 &  & & &  & 7\\
45 & 2.860 [3] & 2.860 $ \pm $ 0.057 & [3]&  & &$<$1 \\
51 & 0.72 $ \pm $ 0.42 &  &&  &  & 5\\
52 & 42.3 $ \pm $ 8.0 & 11.39 $ \pm $ 0.79 & [1] & 17.24 $ \pm $ 1.0 & 11.170 $ \pm $ 8.400 & 10\\
60 & 0.402 $ \pm $ 0.37 &  &&  &  & 6\\
63 & 2.022 $ \pm $ 1.68 &  &&  &  & 6\\
65 & 7.2 $ \pm $ 4.2 & 5.30 $ \pm $ 0.96 & [1] &  & & 5 \\
78 & 3.23 $ \pm $ 2.3 &  &&  &  &9\\
94 & 15.8 $ \pm $ 11.5 &  &&  &  & 7\\
105 & 1.109 [*] &  &&  &  &$<$1\\
107 & 18.2 $ \pm $ 4.6 & 5.630 $ \pm $ 0.169 & [3] &  & &5 \\
130 & 11.1 $ \pm $ 8.0 & 3.320 $ \pm $ 0.199 & [4]&  &  &$<$1\\
135 & 0.917 $ \pm $ 0.88 &  & &&  &2\\
139 & 5.9 $ \pm $ 3.3 &  & & 3.59 $ \pm $ 0.10 & & 17 \\
145 & 2.266 [*] &  &&  &  & $<$1\\
187 & 2.5 $ \pm $ 1.07 & & & & & $<$1  \\
192 & 0.719 [*] & & & 1.37 $ \pm $ 0.41 & &11  \\
194 & 8.8 $ \pm $ 2.9 & & &  &  &5\\
216 & 0.560$ \pm $ 0.46 &  && 3.53 $ \pm $ 0.12 & &3 \\
253 & 0.904 $ \pm $ 0.65 & 0.052 $ \pm $ 0.002 & [5] & & &$<$1 \\
337 & 0.543 $ \pm $ 0.080 & & &  &  & 2\\
344 & 0.340 $ \pm $ 0.19 & & & &  & 7\\
354 & 2.451 [*] & & & 4.88 $\pm$ 0.35 & &10 \\
372 & 4.443 [*] & & &  &  &$<$1\\
419 & 0.997 $ \pm $ 0.55 & & &  & &10 \\
451 & 21.0 $ \pm $ 14.8 & 10.2 $\pm$ 3.4 & [6] &  & &5 \\
488 & 6.2 $ \pm $ 5.5 & & & &  &9\\
511 & 19.9 $ \pm $ 4.1 & 18.96 $ \pm $ 0.99 & [1] &  & 8.580 $ \pm $ 5.930 &10\\
532 & 2.89 $ \pm $ 0.76 & 16.8 $\pm$ 2.8 & [6] & 5.46 $ \pm $ 0.10 & 4.970 $ \pm $ 2.810 &34\\
554 & 1.6 $ \pm $ 1.3 &  && &  & 5\\
704 & 18.600 [2] & 19.65 $ \pm $ 0.89 & [1] & 16.23 $ \pm $ 1.0 & 19.970 $ \pm $ 6.570 &34\\
747 & 6.0 $ \pm $ 2.3 & & & 0.04 $ \pm $ 0.02 &  & 16\\
804 & 2.5 $ \pm $ 1.8 & 1.75 $ \pm $ 0.40 & [1] & &  & 2\\
\hline
\end{tabular}

\label{comparmass}
\end{center}
\end{table}

\section{INPOP10a General features}

\begin{table}
\caption{Values of parameters obtained in the fit of INPOP08 and INPOP10a to observations. The (F) indicates that the marked values are fixed in the fit. The equivalent value of AU deduced from the estimation of the GM$_{\odot}$ in INPOP10a and (Konopliv et al. 2011) \cite{2011Icar..211..401K} are given in the line labelled "AU from GM${\odot}$". For the AU, the GM${\odot}$ and the asteroid masses, the uncertainties are deduced from the least squares fit. }
\begin{center}
\begin{tabular}{l c c c}
\hline
& INPOP08 & INPOP10a & K11\\
&  $\pm$ 1$\sigma$&  $\pm$ 1$\sigma$ & \\
\hline
(EMRAT-81.3000)$\times$ 10$^{-4}$ & (5.4 $\pm$ 0.5) & (5.7 $\pm$ 0.010) & (5.694 $\pm$ 0.015) \\
%$\Delta GM {\odot} / GM{\odot} $ &  & $( \pm ) \times 10^{-10}$ \\
\\
J$_{2}$$^{\odot}$ $\times$ 10$^{-7}$ & (1.82 $\pm$ 0.47)  & (2.40 $\pm$ 0.25)  &\\
\\
%\\
%$\dot{\varpi}$ Mer & (-10 $\pm$ 30) mas$/$yr & & & \\
%$\dot{\varpi}$ Sat & (-10 $\pm$ 8) mas$/$yr & & & \\
\hline
\\
GM$_{\odot}$ - GM$_{\odot}^{DE405}$ [km$^{3}.$ s$^{-2}$]&  0.0 $\pm$ 50 (F) & 37.013 $\pm$ 1 & 24.013 $\pm$ 10 \\
 \\
AU-AU$^{IERS03}$ [m] & 8.2 $\pm$ 0.11 & 0.0 (F) & \\ 
\\
AU [m] from GM${\odot}$ & & 13.9 $\pm$ 0.3 & 9.0 $\pm$ 3 \\
\hline
\end{tabular}
%\hline
\end{center}
\label{paramfita}
\end{table}

\begin{table}
\caption{Values of parameters obtained in the fit of INPOP08 and INPOP10a to observations. The given values are $\beta$ and $\gamma$ intervals in which the differences of postfit residuals from INPOP10a are below 5\%. K11 stands for (Konopliv et al. 2011) \cite{2011Icar..211..401K}, M08 for (Muller et al. 2008) \cite{2008JGeod..82..133M}, W09 for (Williams et al. 2009) \cite{2009arXiv0901.0507083}, B03 for (Bertotti et al. 2003) \cite{2003Natur.425..374B} and L09 for (Lambert and al. 2009) \cite{2009A&A...499..331L}}
\begin{tabular}{l l c c }
\hline
Fixed PPN & Estimated PPN & & $\times$ 10$^{4}$\\
% &  & &  \\
\hline
\\
$\gamma$  = 1 & ($\beta$ -1)  & INPOP08 & (0.75 $\pm$ 1.25) \\
& & INPOP10a & (-0.62 $\pm$ 0.81) \\
\\
$\gamma$ = $\gamma_{B03}$ & ($\beta$ -1) & INPOP10a & (-0.41 $\pm$ 0.78) \\
& & K11 & (0.4 $\pm$ 2.4) \\
& & M08-LLR-SEP & (1.5 $\pm$ 1.8) \\
& & W09-LLR-SEP & (1.2 $\pm$ 1.1) \\
\\
$\beta$  = 1 & ($\gamma$ -1) & INPOP10a &  (0.45 $\pm$ 0.75) \\
& & K11 &  (1.8 $\pm$ 2.6) \\
& & B03-CASS & (0.21 $\pm$ 0.23) \\
& & L09-VLB & (0.7 $\pm$ 2.0 )\\
\hline
\end{tabular}
\label{paramfitc}
\end{table}

%\vspace{0.5cm}
\begin{table}
\caption{Values of parameters obtained in the fit of INPOP08 and INPOP10a to observations. The supplementary advances of perihelia and nodes are estimated in INPOP10a and INPOP08 as the interval in which the differences of postfit residuals from INPOP10a are below 5\%. P09 stands for (Pitjeva 2010 \cite{2010IAUS..261..170P}).}
\begin{tabular}{l l l l}
\hline
  &INPOP08 & INPOP10a & P09\\
\hline
$\dot{\varpi}_{\mathrm{sup}}$ &  & & \\
mas $\times$ cy$^{-1}$ &  & & \\
\\
Mercury & -10 $\pm$ 30 & 0.4 $\pm$ 0.6 & -4 $\pm$ 5 \\
Venus & -4 $\pm$ 6& 0.2 $\pm$ 1.5 & 24 $\pm$ 33\\
EMB & 0.0 $\pm$ 0.2 & -0.2 $\pm$ 0.9 & 6 $\pm$ 7  \\
Mars & 0.4 $\pm$ 0.6 & -0.04 $\pm$ 0.15 & -7 $\pm$ 7  \\
Jupiter & 142 $\pm$ 156 & -41 $\pm$ 42 & 67 $\pm$ 93 \\
Saturn & -10 $\pm$  8 & 0.15$\pm$ 0.65 & -10 $\pm$  15 \\ 
\\
$\dot{\Omega}_{\mathrm{sup}}$  & & & \\
mas $\times$ cy$^{-1}$&  & & \\
\\
Mercury & &  1.4 $\pm$ 1.8 & \\
Venus & 200 $\pm$ 100 & 0.2 $\pm$ 1.5 & \\
EMB & 0.0 $\pm$ 10.0 & 0.0 $\pm$ 0.9  & \\
Mars & 0.0 $\pm$ 2 & -0.05 $\pm$ 0.13& \\
Jupiter & -200 $\pm$ 100 & -40 $\pm$ 42 & \\
Saturn & -200 $\pm$  100 & -0.1 $\pm$ 0.4  & \\
\hline
\end{tabular}
\footnotetext[1]{fixed}
\label{paramfitd}
\end{table}

\smallskip

In INPOP10a, the motion of the planets including Pluto and the Moon are integrated with the rotations of the  Earth and the Moon, the (TT-TDB) differences and the orbits of several hundreds asteroids as in INPOP08 (Fienga et al. 2009 \cite{2009A&A...507.1675F}).
The initial conditions of the planet orbits are fitted over the data sets described in section \ref{section2}. The Moon orbit and rotation initial conditions are fitted separately by Manche et al. (2010) \cite{Manche2010}. An iterative process between planetary and lunar adjustments is used to keep consistencies between the two part of the ephemerides. 
With the initial conditions of the planets are also fitted the  Earth-Moon mass ratio and the oblateness of the Sun coefficient $J_{2}$.

In opposition to INPOP08 where the value of astronomical unit (AU) is fitted, in INPOP10a, the value of the AU is fixed to the value given by the IERS2003 conventions \cite{McCarthyIERS32}. 
The GM of the Sun is fitted to the observations like the other fitted quantities. Values of the fitted parameters are given in table \ref{paramfita}.
(Konopliv et al. 2011) \cite{2011Icar..211..401K} presents also a value of the GM of the Sun.
The two estimations are consistent at 2 sigmas with a ten times smaller uncertainty for the INPOP10a estimation.
This can be explained by the fact that the INPOP10a uncertainties are 1 formal sigma deduced from least square fit when the uncertainties given by (Konopliv et al. 2011) \cite{2011Icar..211..401K} also involved the propagation of the error induced by the use of fixed asteroid masses in the fit.

%We used the mass of planets provided by the IAU 2009 Current Best Estimates lists (Luzum 2009 \cite{2009IAU...261.0301L}).

\subsection{The fit procedure for the asteroid masses}

The estimations of the asteroid perturbations on planet orbits are a critical point for the extrapolation capabilities of the planetary ephemerides (Standish and Fienga 2002 \cite{2002A&A...384..322S}).

The usual approach to this problem has been suggested by Williams 1984 and consists
of including for a selection of approximately 300
individual asteroids to the dynamical model. The masses of the most perturbing asteroids are fitted
to observations. For the other objects, masses are deduced from
radiometric diameters and the assumption of constant densities within three
taxonomic classes. This classic approach has been used in INPOP08 and
achieves in terms of the Earth-Mars distance prediction  an accuracy of 20
m over 2 years. It is based on an
unrealistic hypothesis of constant densities within taxonomic classes.
It also relies on an empirical choice of the selection of asteroids to
account for and on the choice of the subset of asteroid masses to adjust
individually. 
We used INPOP10a as a benchmark to test an alternative approach. Kuchynka
et al. (2010) \cite{2010A&A...514A..96K} showed that approximately 240 asteroids in a list of 287 probable asteroids and a ring should
represent the perturbations induced by the main belt on planetary orbits
down to an order of a meter. The result was obtained for various test
models of the main belt containing each several thousands of asteroids with
randomly assigned masses.
In INPOP10, we use the Bounded Variable Least Squares (BVLS)
algorithm developed by (Lawson and Hanson 1995 \cite{lawson:hanson:1995}) and (Stark and Parker 1995 \cite{Stark&Parker95}) in order to fit the masses of all
the 287 asteroids listed in Kuchynka et al. (2010) \cite{2010A&A...514A..96K}. The BVLS solves the
least square problem with constraints which require the adjusted masses to
be positive or zero. Setting an asteroid mass to zero is equivalent to
removing it from the dynamical model. Thus the BVLS algorithm performs
simultaneously parameter selection and estimation. 
From this method and the original list of 287 asteroids, about 161 asteroid masses were planed to be estimated in the fit, the other masses being put to zero.
 
Based on a study of the correlations between the asteroid masses, we found however 30 highly correlated asteroids among the most perturbing objects.
In order to decrease the mass uncertainties, we fixed 16 asteroid masses to values 
 determined by other methods (close encounters, binary) or by previous published or unpublished INPOP versions as INPOP08 or INPOP08e.
in table 2, the fixed masses are identified with their references.

Besides these fixed values, we then estimate 145 asteroid masses using constraints on the densities  in order to keep the fitted values in the frame of realistic physics. These densities are constrained to be positive and smaller than 20 g$\times$cm$^{-3}$. The diameters used to estimate the densities
are extracted from a database (Kuchynka 2010) \cite{KuchynkaPHD} compiling mainly diameters from the IRAS catalogue (Tedesco et al. 2002 \cite{tedesco2002}).  For objects without radiometric data, diameters are estimated from assumed mean albedo (Tedesco et al. 2005 \cite{tedesco2005}).

%\begin{figure}
%\includegraphics[scale=0.5]{inpop10a_421_RDD_4b2.jpeg}
%\includegraphics[scale=0.5]{inpop10a_421_RDD_4z2.jpeg}
%\caption{Comparisons of geocentric right ascensions, declinations and distances of Mars based on  INPOP10a, DE421, INPOP08 and INPOP06}
%\label{mm}
%\end{figure}

%\begin{figure}
%\includegraphics[scale=0.5]{inpop10a_421_RDD_9b.jpeg}
%\caption{Comparisons of geocentric right ascensions, declinations and distances of Pluto based on  INPOP10a, DE421, INPOP08 and INPOP06}
%\label{plu}
%\end{figure}

\section{ Results and applications}
\label{results}

\subsection{Postfit residuals and comparisons with other planetary ephemerides}

%In figures 1 to 6 of the Online Resource 1 are plotted the differences between INPOP10a and DE421 in right ascension, declination and geocentric distance for Mercury to Pluto. We also plot the differences obtained for Jupiter between INPOP10a, INPOP08, INPOP06 and DE421.
%Differences between the same ephemerides obtained for the  Earth-Moon barycenter orbit versus the barycenter of the Solar System are given in figure \ref{eb} in longitude, latitude and barycentric distance.
Complete plots of comparisons between INPOP06, INPOP08, INPOP10a and DE421 can be find in Online Resource 1 (Figures 1 to 6).
For all the planets, one can note the convergence of the last INPOP solution and DE421. 
%Complete plots of comparisons between INPOP06, INPOP08, INPOP10a and DE421 can be find in Online Resource 1 (Figures 1 to 6). %On these plots, we see that we obtain with INPOP10a an important improvement of  Jupiter orbit compared to INPOP08. 
We stress in particular the reduction of the differences between planetary ephemerides for the Jupiter orbit. As one can see in figure 4, INPOP08 appears then to be biased especially for Jupiter orbit. This can be explained by the fact that no normal points deduced from mission flybys were included in INPOP08 while a large amount of Mars mission data were presented. The INPOP08 estimated orbit of Jupiter seems then to be over-weighted by Mars data. In the case of INPOP06, no flyby points were included but the fit was far less Mars observation-depend than INPOP08. In INPOP10a, Mars observations are still very important but the input of the flyby normal points balances the impact for Jupiter's orbit. Furthermore, due to transmission problems during the mission, VLBI Galileo angular positions have a degraded accuracy compared to the one expected . Thus, even if these points were included in the INPOP08 adjustment, they were not reliable enough  to correct Jupiter's orbit. 
For Venus, the input of the VEX data is obvious in figure 1 as for Saturn and the impact of Cassini data.
The noticeable offsets for Saturn and Jupiter longitudes between INPOP10a and DE421 are far below the limit of accuracy of the data used for the fit. 
For Uranus and Neptune (figure 5), the differences between the ephemerides are about the same with a change of behavior induced by the input of the flyby points in 1986 for Uranus and 1989 for Neptune. 
For Mercury, despite the input of the five flyby points, the differences between the ephemerides including or not these points are about the same.  

Table \ref{omctab} gives the planetary postfit residuals obtained with INPOP08 and INPOP10a.
Some of the data sets used in INPOP10a were not used in the INPOP08 adjustment: this explains some of the differences between the two columns of Table \ref{omctab}.
Figures 1 to 6 in Online Resource 2 and Figures 1 to 3 in Online Resource 3 present the postfit residuals obtained with INPOP10a for the different sets of observations used in the fit. 
%As one can see on these plots, INPOP08 shows a good improvement of the inner planet orbits thanks to the input of VEX and MEX data. But this improvement comes together with a degradation of the outer planet orbits 
As one can see with the extrapolated residuals of INPOP08 for the outer planet flybys (see Table \ref{omctab}), the improvement of the inner planets came together with a degradation of the outer planet orbits, no noticeable with the postfit residuals obtained at that time (Fienga et al. 2009 \cite{2009A&A...507.1675F}). In other words, the outer planet orbits of INPOP08 are accurate at the level of uncertainty of the optical and VLBI observations. INPOP10a, with the use of the flyby data of the outer planets and the improvement of the asteroid masses, provides good residuals for all the planets. Figures of INPOP06 and INPOP08 residuals obtained with the INPOP10a data set are given in Online Resources 2 and 3.

\begin{figure}
 \begin{center}
 \includegraphics[width=12cm]{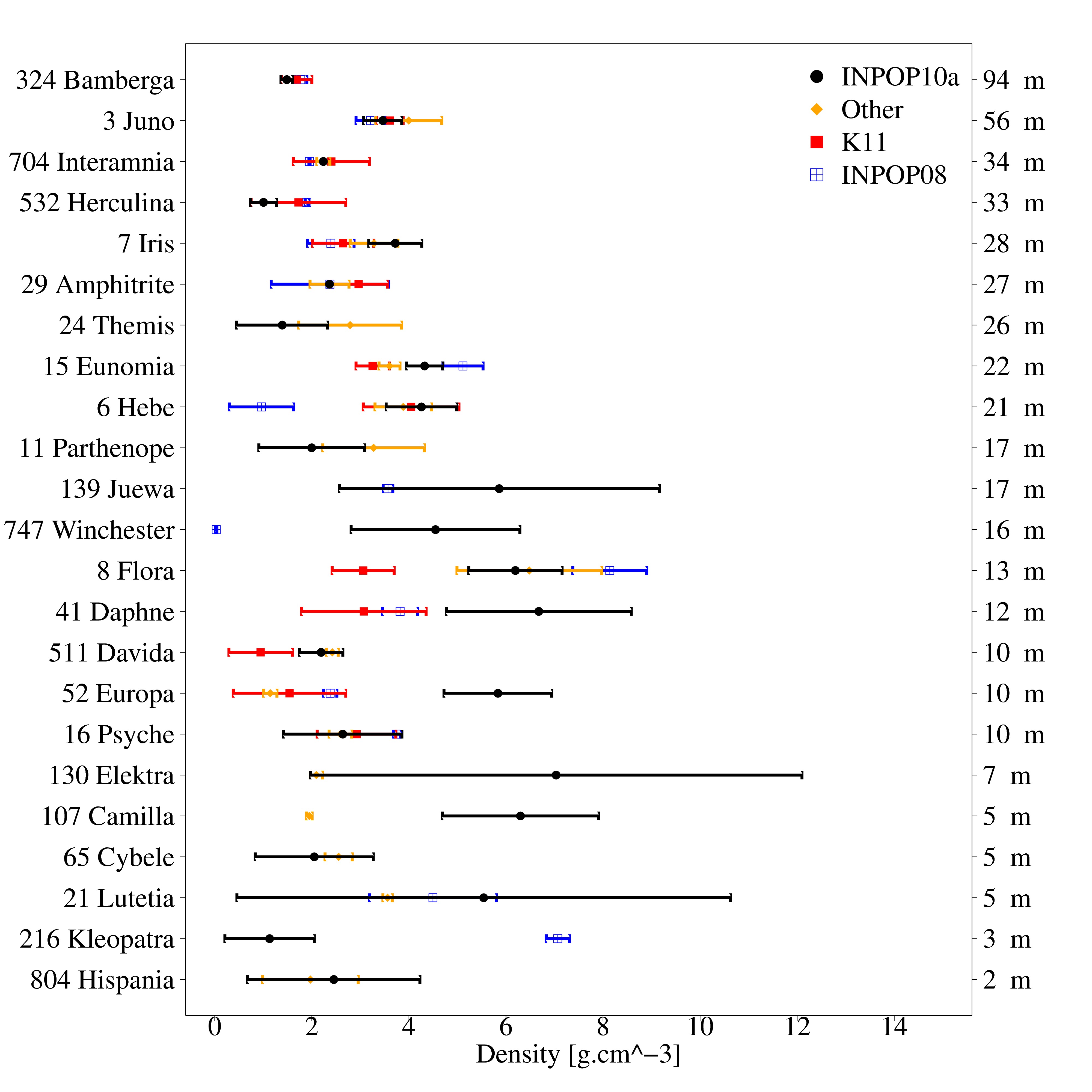}
  \end{center}
%\vspace{-2cm}
\caption{Distribution of densities in g $\times$ cm$^{-3}$ ranked with the impact on the Mars- Earth distances over 1970-2010 period. The right-hand side axis gives the differences in Mars geocentric distances in meters induced by an integration of the Mars motion over the 1970-2010 period with and without the corresponding asteroid. {\it{K11}} stands for (Konopliv et al. 2011) \cite{2011Icar..211..401K} and {\it{Other}} for the determinations induced by close-encounters or binary systems. The error bars represent the 1-sigma error on the mass determinations.}
\label{compmass}
 \end{figure}

\subsection{Asteroid masses}

in figure \ref{compmass} are presented the asteroid densities deduced from INPOP10a compared to values found in the literature, ranked 
 by their impact on the  Earth-Mars distances over the 1970 to 2010 period. The major sources of comparisons are the values obtained with INPOP08, (Konopliv et al. 2011) \cite{2011Icar..211..401K} and 
close encounters or binary system estimations gathered by (Baer et al. 2011 \cite{2011Icar..212..438B}). 
The densities are deduced from the mass estimations and diameters extracted from the (Kuchynka 2010) database. 

\begin{figure}
 \begin{center}
\includegraphics[scale=0.25]{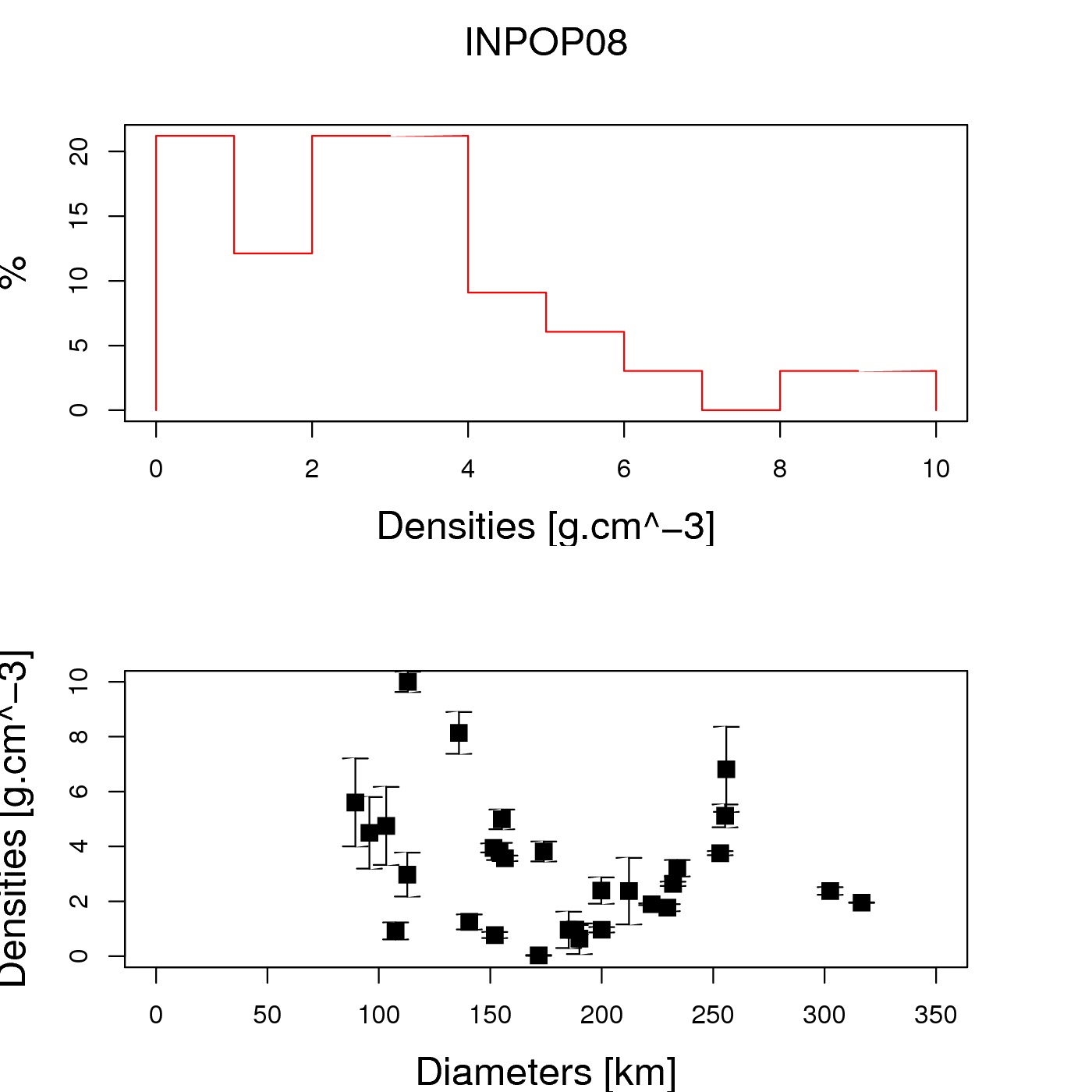}\includegraphics[scale=0.25]{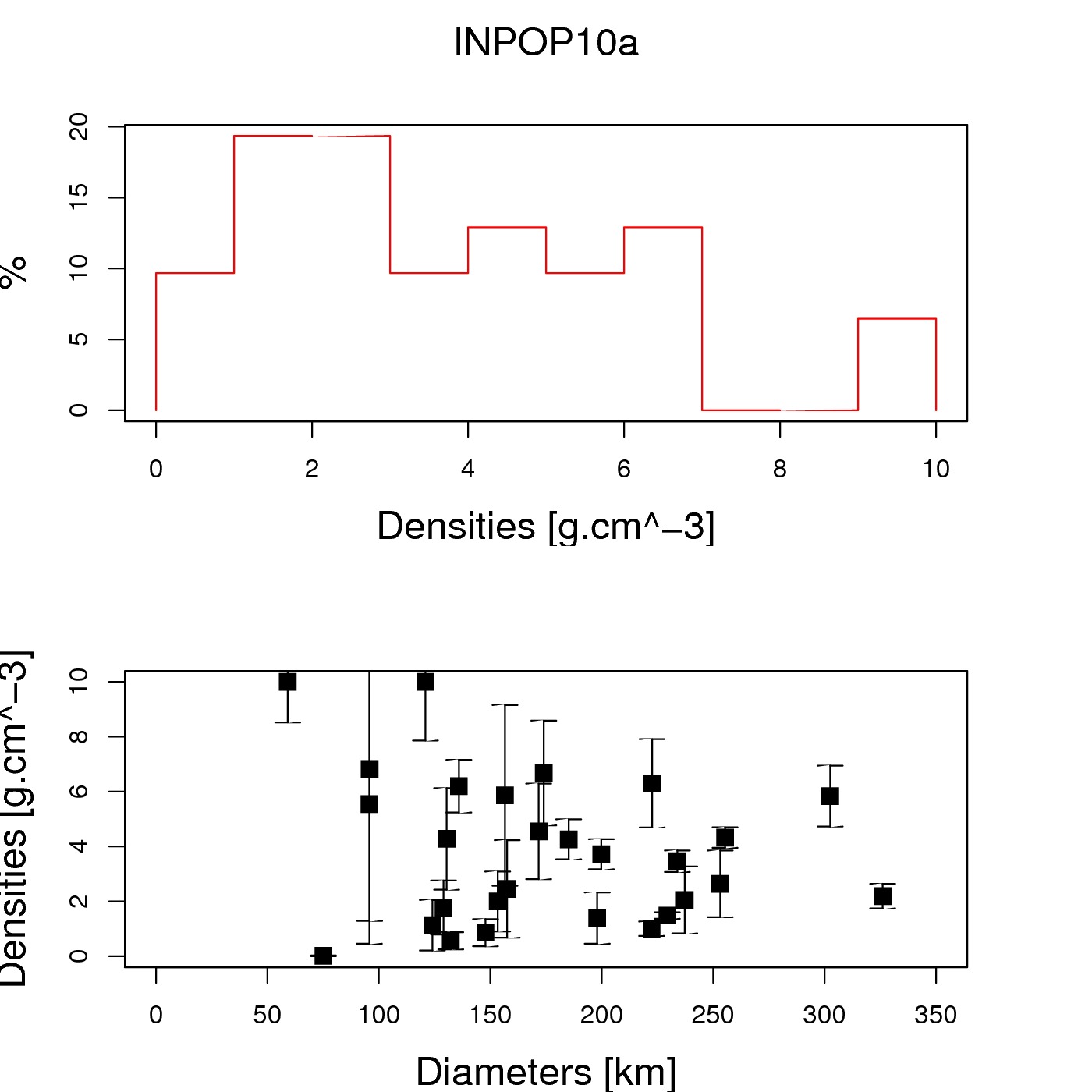}\includegraphics[scale=0.25]{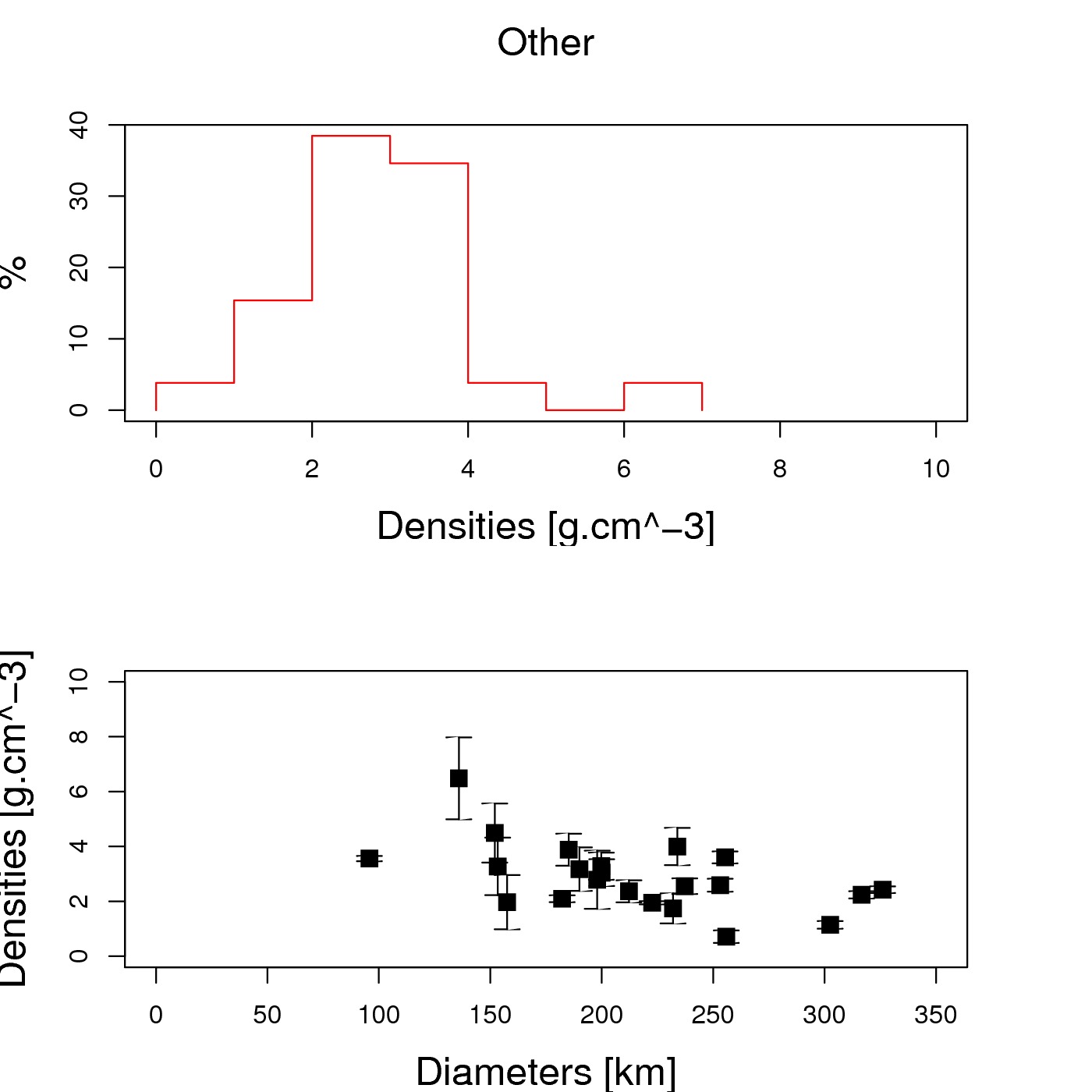}
\end{center}
\caption{Histograms of densities in g $\times$ cm$^{-3}$ for the most perturbing objects deduced from INPOP10a, INPOP08, and Other determinations (close encounters and binary system) and distributions of densities versus the diameters in kilometers. The plotted uncertainties represent the 1-sigma error on the mass determinations.}
%\vspace{-10cm}
\label{histdens}
 \end{figure}

It appears clearly from figure \ref{compmass} that the estimations for the most perturbing objects are quite consistent while the estimations of the less perturbing objects show bigger discrepancies. 

in tables \ref{comparmass} are given the details of asteroid masses obtained with the adjustment of INPOP10a. As expected the three bigger asteroid masses are compatible at 2-sigmas. 
%On can note the high improvement of the uncertainties on the masses of Pallas and Vesta between INPOP08 and INPOP10a, mainly induced by the change of the fitting procedure.
 
in figure \ref{histdens} are plotted the distributions of densities deduced from INPOP10a, INPOP08 and from close encounters and binary systems (noted "Other"). Two representations are given: one histogram of density distribution and one distribution of the density versus the diameters of the objects.
Are plotted on these figures, only the densities deduced from perturbations bigger than 1 meter on the Mars-Earth distances over the 1970 to 2010 period with error bars representing the 1-sigma uncertainties on the mass determination. The diameters are considered here as perfect.
With this optimistic hypothesis, one can first note the  smaller uncertainties on the close-encounter estimations compared to those obtained with INPOP08 and INPOP10a. The distribution of the close-encounter densities shows also two very close peaks of frequency at 2.5 and 3.5 g $\times$ cm$^{-3}$, and about few percentages of objects with a density below 1 g $\times$ cm$^{-3}$. This trend is opposite to the INPOP08 distribution with more than 20 \% of low density asteroids.
 This behavior was noticed in (Fienga et al. 2009 \cite{2009A&A...507.1675F}) and shows the excess of under-estimated masses in INPOP08. 
In the other hand, the dispersion and the uncertainties of the INPOP10a distribution do not allow to give conclusive remarks even if one can note a diminution of the number of  low density objects compared to INPOP08.

\subsection{Gravity tests}

\begin{figure}
\begin{center}
\includegraphics[width=6cm]{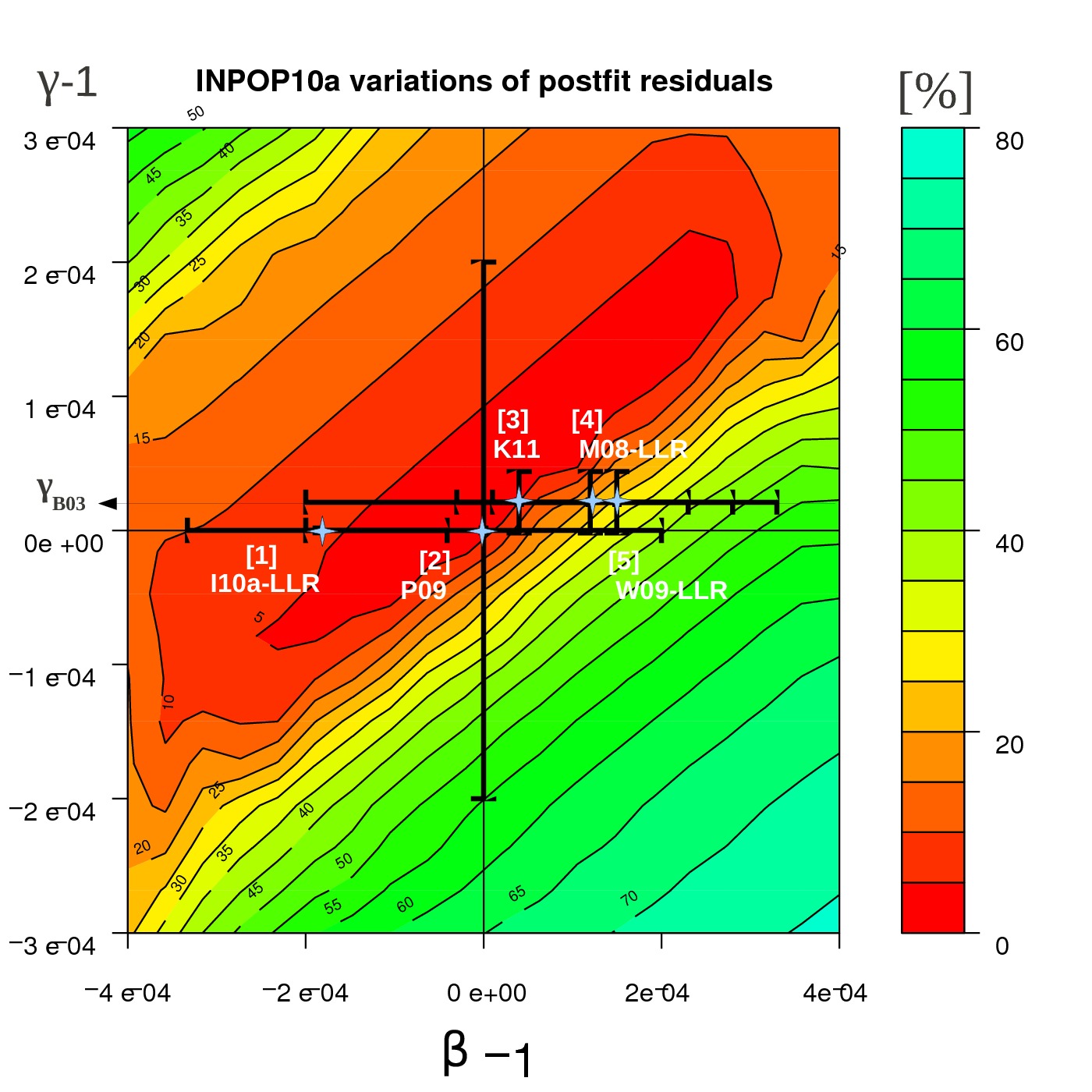}\includegraphics[width=6cm]{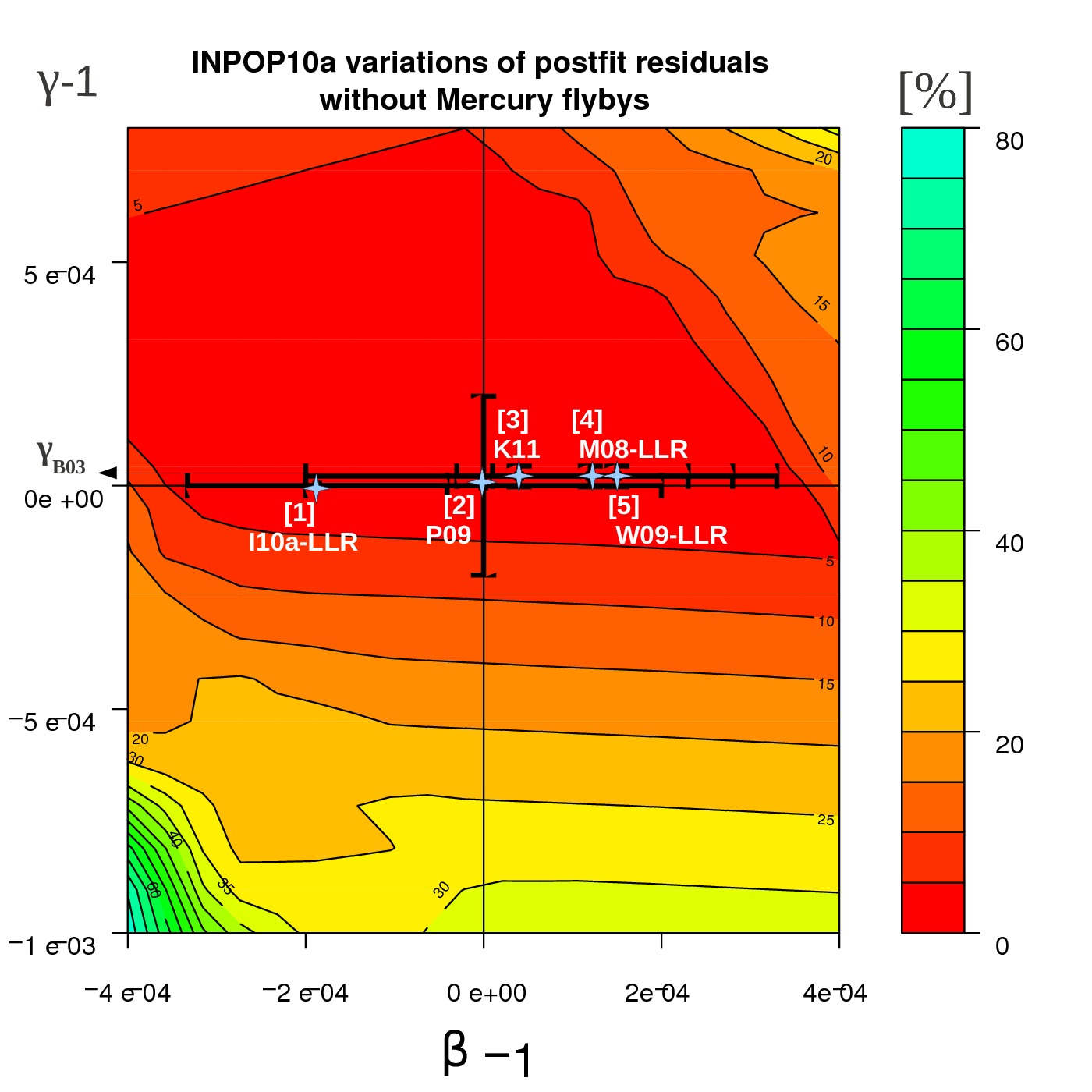}
\end{center}
\caption{Variations of postfit residuals obtained for different values of PPN $\beta$ (x-axis) and $\gamma$ (y-axis). [1] stands for a PPN $\beta$ value obtained by (Manche et al 2010 \cite{Manche2010}) using LLR observations with $\gamma=0$, [2] stands for Pitjeva (2009) \cite{2010IAUS..261..170P} by a global fit of EPM planetary ephemerides. K11 stands for (Konopliv et al. 2011) \cite{2011Icar..211..401K} determinations based mainly on Mars data analysis. M08 for (Muller et al. 2008) \cite{2008JGeod..82..133M} and W09 for (Williams et al. 2009)  \cite{2009arXiv0901.0507083} give values deduced from LLR for a fixed value of $\gamma$, B03 stands for (Bertotti et al. 2003) \cite{2003Natur.425..374B} determination of $\gamma$ by solar conjunction during the Cassini mission.}
\label{betagamma}
\end{figure}

 Table \ref{paramfita} gives the obtained values for the GM of the Sun, the Sun $J_{2}$, the  Earth-moon mass ratio and table \ref{paramfitc} the interval of
sensitivity of data to modifications of PPN $\beta$ with $\gamma$ equal to 1 or with $\gamma$ equal to the value obtained by the Cassini experiment (Bertotti et al. 2003 \cite{2003Natur.425..374B}), and of PPN $\gamma$ with $\beta$ equal to 1. The results obtained for both modified parameters are presented in figure \ref{betagamma}.
Supplementary advances in the perihelia and nodes of the planets (from Mercury to Saturn)  have also been tested and are presented in table \ref{paramfitd}.
All these results were  based on the method presented in Fienga et al. (2010) \cite{2010IAUS..261..159F}. 
We computed several fits for different values of the PPN parameters ($\beta$, $\gamma$) or supplementary advances in perihelia and nodes with a simultaneous fit of initial conditions of planets, mass of the Sun and asteroid densities.
The given intervals correspond to values of parameters inducing changes in the postfit residuals below 5$\%$ compared to INPOP10a postfit residuals. 
in figure \ref{betagamma} are plotted the variations of postfit residuals induced by the modification of the corresponding $\beta$ and $\gamma$ parameters. The left hand side plot gives the variations of postfit
 residuals including Mercury flyby normal point when the right hand side plot gives the variations of residuals without the Mercury observations.
%As one can see in table \ref{paramfitc}, the obtained PPN intervals are compatible with each other with a better accuracy of INPOP10a due to the use of Mercury normal points. 

The different levels of grey indicate the percentage of variations of the postfit residuals compared to those obtained with INPOP10a. By projecting the 5\% area on the $\beta$-axis (or the $\gamma$-axis), one can deduced the corresponding $\beta$ (or $\gamma$) interval in which the residuals are modified by less than 5\%.
In looking at the two figures, one can see that the use of the Mercury flyby data give smallest intervals of possible $\beta$,$\gamma$. This is consistent with the fact that the Mercury observations are far more sensitive to gravity modifications than other data (see table 1 in (Fienga et al. 2010) \cite{2010IAUS..261..159F} ).

As one can see in table \ref{paramfitc} and in figure \ref{betagamma}, the possible variations acceptable for all the residuals including the Mercury points are compatible with the estimations of $\beta$ deduced from planetary ephemerides as (Konopliv et al. 2011)\cite{2011Icar..211..401K} and (Pitjeva 2009)\cite{2010IAUS..261..170P}. However, one can see that the $\beta$ determinations constrained by Mercury flybys are not consistent with the one deduced from LLR measurements of the equivalence principle ((Muller et al. 2008) \cite{2008JGeod..82..133M} and (Williams et al. 2009)  \cite{2009arXiv0901.0507083}). If the Mercury flyby data are not taken into account (figure \ref{paramfitc}, right hand side plot), the constraint is then released and all the determinations become then compatible. 
The Messenger spacecraft is now orbiting Mercury. The use of its tracking data will help to improve the $\beta$, $\gamma$ estimations and to confirm or not these denoted discrepancies.
For the advances of perihelia and the nodes, the estimations based on INPOP10a show the clear incompatibility of a significant supplementary advance in the planets' orbits and the observations used in the INPOP10a adjustment. This conclusion is due to the densification of very accurate Cassini observations around Saturn. 

%\begin{figure}
%\begin{center}
%\includegraphics[width=10cm]{journees_26.jpeg}
%\end{center}
%\caption{Variations of postfit residuals obtained for different values of PPN $\beta$ (x-axis) and $%\gamma$ (y-axis). The lines }
%\label{betagammaA}
%\end{figure}

\subsection{Use of millisecond pulsars for reference frame ties}

One possible method to test the real accuracy of planetary ephemeris is to use very accurate observations of objects but not used in the fit process of the ephemerides. 
%We can then test the quality of the new planetary ephemerides by using the new  Earth orbit in the reduction procedures of such observations. The deduced positions are then obtained in the reference frame of the new ephemerides. If the same object is also directly observed in ICRF, we can also test the link between the planetary ephemerides frame and ICRF.
Nowadays three types of observations have a mas level astrometry: the VLBI surveys of ICRF sources, the VLBI tracking of spacecrafts into ICRF used in the construction of the planetary ephemerides and the millisecond pulsar (MSP) observations by radio timing and by VLBI  astrometry relative to ICRF sources. 
%In this context of planetary ephemerides testing, these last type of pulsar observations appear to be a good opportunity.
Pulsar timing is the regular monitoring of the rotation of the neutron star by tracking (nearly exactly) the times of arrival of the radio pulses. 
By their relatively stable rotations and sharp pulses of nearly point-source radio emission, millisecond pulsars can have the astrometry accurate at the mas level if series of observations are done on a long enough time interval. 

Every single rotation of the neutron star is unambiguously accounted over long period of time.
This unambiguous and very precise tracking of rotational phase allows pulsar astronomers to probe the physics interior  of neutron stars, to make extremely accurate astrometric measurements, and test gravitational theories in the strong-field regime in unique ways.
Of course, systematics in the pulse observations exist such as the fluctuations of the interstellar medium (DM) or the binary structure of the neutron star. 
 However, by having a long timespan of observations it becomes possible to model the frequencies induced by the companion or to average the DM effect and then reduce significantly the error.
Multi wavelength observations of millisecond pulsars are also performed in order to study the physics and the dynamics of these objects as the VLA polarization survey done by (Taylor et al. 1993 \cite{1993ApJS...88..529T}) and (Han and Tian 1999 \cite{1999A&AS..136..571H}) or the several year program using VLBA and VLBI techniques begun by (Chatterjee et al 2001 \cite{2001ApJ...550..287C}) to estimate proper motions and parallaxes of about 100 pulsars.
The mas level astrometry of pulsars are then an interesting tool for testing planetary ephemerides.
 By the combination of  millisecond pulsar timing and VLBI observations of the same objects, it is possible to establish a new independent link between ICRF and dynamical reference frames.
 (Shapiro and Knight 1970 \cite{1970ASSL...20..284S}) have the first proposed to establish such direct link.

Since 2009, VLBI observations of MSP are regularly done by (Chatterjee et al. 2009 \cite{2009ApJ...698..250C}) in the northern hemisphere and (Deller et al. 2009 \cite{2009ApJ...690..198D}) in southern hemisphere. 
These two surveys provide positions of MSP in the ICRF at the mas level and give then a good opportunity to realize the link between ICRF and planetary ephemerides reference frames.

VLBI-derived positions of MSP can provide a tie between the extragalactic ( Earth-rotation based) reference
frame in which VLBI operates and the dynamic ( Earth-orbit based) reference frame in which MSP timing
positions are derived. 

If ($\alpha_{PE1}$,$\delta_{PE1}$) are positions of a MSP deduced from radio timing in using Planetary Ephemerides 1 (PE1) and ($\alpha_{PE2}$,$\delta_{PE2}$) are positions of the same MSP deduced either from VLBI observations in ICRF either from radio timing in using Planetary Ephemerides 2, the differences($\alpha_{PE1}$,$\delta_{PE1}$) and ($\alpha_{PE2}$,$\delta_{PE2}$) can be seen as residual rotations $R_{x}(\theta)$,$R_{y}(\eta)$ and $R_{z}(\zeta)$ about the x,y and z axis of PE1 reference frame  such as 
\begin{equation}
\alpha_{PE1} = R_{x}(\theta)R_{y}(\eta)R_{z}(\zeta) \alpha_{PE2} 
\end{equation}
and 
\begin{equation}
\delta_{PE1} = R_{x}(\theta)R_{y}(\eta)R_{z}(\zeta) \delta_{PE2}
\end{equation}.
The angles $\theta$,$\eta$ and $\zeta$ can then be deduced by least square fitting.

From the radio timing profiles obtained at the NRT (Desvignes 2009 \cite{Desvignes2009}), we selected 18 MSP with radio timing astrometry better than 10 mas. These 18 MSP presented in table \ref{psr} are used, as a check of the procedure, for the estimation of the rotation matrices between DE421 (Folkner et al. 2008 \cite{DE421}), INPOP08 (Fienga et al. 2009 \cite{2009A&A...507.1675F}), DE405 (Standish 1998 \cite{DE405}) and DE200 (Standish 1990 \cite{1990A&A...233..252S})  presented in table \ref{ICRFangles}.
For the second step, we collected in the literature positions and velocities of millisecond pulsars obtained by VLBI astrometry (Chatterjee et al. 2009 \cite{2009ApJ...698..250C}, Deller et al. 2009 \cite{2009ApJ...690..198D}) and observed by the NRT.
4 MSP have a mas-level accuracy in both techniques,VLBI and radio timing, and  are used to test the link between planetary ephemerides frames and ICRF. Results are presented in table \ref{ICRFangles}

 \begin{table}
\caption{ Main characteristics of millisecond pulsars used for this study. The last column indicates by a $\times$ if the pulsar is used for the ICRF link. \cite{2009ApJ...698..250C} stands for (Chatterjee et al. 2009), \cite{2009ApJ...690..198D} for (Deller et al. 2009), \cite{1998ASPC..144..331N} for (Nunes and Bartel 1998), \cite{1996AJ....112.1690B} for (Bartel et al. 2006), \cite{Petit94} for (Petit 1994). }
\begin{center}
%\include{table1.txt}
%\tiny
\begin{tabular}{r | c r r l | c c l | c}
\hline
&\multicolumn{4}{c}{Radio Timing} & \multicolumn{2}{c}{VLBI}& \\
MSP & RMS & $\sigma_{\alpha}$ & $\sigma_{\delta}$ & Period &  Ref & $\sigma$ & &\\
& $\mu$s & mas& mas & &  & mas &\\
\hline

J0139+58 &  62 & 800& 70 & 2005-2009.5&  \cite{2009ApJ...698..250C} & 15 &\\
J0454+55 &  90 & 100 & 100 & 2004.5-2009& \cite{2009ApJ...698..250C} & 2&\\
J0613-02 & 1.6 & 0.6 & 0.15 & 2004.5-2009.5 && &  \\
{J0737-30} & 44 & 0.6 & 0.6 &2005-2009.5 & \cite{2009ApJ...690..198D} & 0.4 & $\times$ \\
J0751+18 & 2.6 & 3 & 14 &2005-2009.5&&&\\
J1012+53 & 2.6 & 3.0 & 2.0 &2005-2009.5&&&\\
J1022+10 & 3.2 & 135 & 360 &2005-2009.5&&&\\
J1024-0719 & 1 & 0.6 & 2 & 2006-2009.5  &&&\\
J1300+12 & 9 & 18 & 36 &2005-2009& \cite{1998ASPC..144..331N} & 30& \\
J1455-33 & 3.2 & 1.5 & 3.0 &2005-2009.5&&&\\ 
J1643-12 & 2.1 & 0.7 & 0.4 &2005-2009.5&&&\\
{J1713+07} & 0.2 & 2& 1&2005-2008.5& \cite{2009ApJ...698..250C} & 2 & $\times$\\
J1730-2304 & 1.3 & 1 & 245 & 2005-2009.5&&& \\
J1744-11 & 0.9 & 0.9 & 0.4 &2005-2009.5&&& \\
J1824-24 &  14.0 & 0.3 & 0.3 &1990-2008&&& \\
J1857+0943 & 0.8 & 1 & 3 & 2005-2009.5&&& \\
B1909-37 &  0.3 & 0.8 & 0.7 &2005-2007&&& \\
%B1937+21 & 2.5 & 0.1& 6 & 1989-2007 &[3],[4] & 3 &$\times$\\
{B1937+21} & 0.3 & 0.9& 1 & 2005-2007 &\cite{1996AJ....112.1690B}, \cite{Petit94} & 3 &$\times$\\
{J2145-07} & 2.1 & 1 & 4 &2005-2009.5& \cite{2009ApJ...690..198D} & 1& $\times$\\

%MSPJ1824-2452 & 7 & 1 & 0.1 & & \\
\hline
\end{tabular}
\end{center}

\label{psr}
\end{table}

We have first estimated the impact of using different planetary ephemerides (with different sets of planet masses) on the analysis of radio timing data in order to check the method. For the 18 MSP used for this study, no specific trend remains in the postfit residuals of the radio timing despite the change in planet masses and initial conditions brought by each different ephemerides.
Among the parameters fitted during the radio timing data analysis, the positions and the proper motions of the pulsars are sightly modified. No other parameters related to the distance, the rotation or the orbit of the companion are affected by the change of planetary ephemerides. 

Based only on radio timing observations, the angles presented in table \ref{ICRFangles} and obtained in using the 18 MSP are quite compatible with previous determinations of rotation matrices between planetary frames. 
As expected, the obtained matrices are consistent with results by (Folkner et al. 1994 \cite{1994A&A...287..279F}) and (Standish 1998 \cite{DE405}).
%This is also consistent with the fact that DE200 was not directly linked to ICRF.
%VLBI tracking data of spacecraft were not used for the DE200 construction  and only optical observations of planets with an accuracy of about 100 mas were used to link the planetary frame to the ICRF. One can also note the improvement of the link based on VLBI tracking data of spacecraft relative to ICRF since DE405. Indeed, there is a significant rotation of about 1.5 mas between DE405 and the later planetary ephemerides DE414, INPOP08 and INPOP10a, these three solutions being very close to each other.
These results confirm the mas-level consistency of the planetary ephemerides and check estimations of the internal accuracy reached for the  Earth orbit. 

in table \ref{ICRFangles} are given the rotation angles obtained between planetary ephemerides frames and ICRF as deduced from pulsar timing and VLBI astrometry. 
Among the 18 selected MSP, only four of them have a mas level accuracy in VLBI and radio timing astrometry. in table  \ref{ICRFangles}, we give the angles between planetary frames only on the four pulsars. These angles are consistent with the ones obtained with 18 objects. It seems then that no peculiar bias is introduced by this selection restricted to only four sources. 
in table \ref{ICRFangles} are also given the rotation angles between the planetary ephemerides and the ICRF. The uncertainties are here about one order of magnitude larger than the internal accuracy estimated previously. One can note a significant rotation of about 10 mas in the y-axis common for all the planetary ephemerides. 
Based on the expected accuracies of the VLBI tracking of spacecraft from 0.5 to 10 mas, the other rotation angles in x and z-axis are not significant. 
It is no clear if the detected rotation in y-axis is real or induced by a the limited number of pulsars used to test the direct link between planetary frames and the ICRF.
An increase of the number of the MSP with a mas level VLBI astrometry will help to validate this conclusion.
We see here the efficiency of using radio timing and VLBI astrometry of millisecond pulsars to estimate the rotation matrices between the planetary frames and the ICRF. We checked that the procedures give similar angles between planetary ephemerides frames as the one already deduced by other methods (Standish 1998 \cite{DE405}, Folkner et al. 1994 \cite{1994A&A...287..279F}) and confirmed the mas level internal accuracy of the planetary ephemerides.
More MSP should be observed in VLBI in order to increase the sample to directly link  planetary frames and ICRF. 

\begin{table}
\caption{Angles of rotation deduced from adjustment of rotation matrices following equations 1 and 2. 
The angles are given in mas and the uncertainties are the formal 1-sigma deduced from the least squares.
In the first part of the table, the angles were obtained in using 18 MSP observed at NRT with an astrometric accuracy better than 10 mas.
In the second part of the table, angles obtained in using 4 MSP (J0737-30, J1713+07, B1937+21, J2145-07) observed in radio timing with an astrometry better than 10 mas are given.
In the third part of the table, ICRF to planetary frame angles are obtained by using the same 4 MSP observed  with an astrometry better than 10 mas both in VLBI and in radio timing.
\cite{1994A&A...287..279F} stands for (Folkner et al. 1994), \cite{DE405} for (Standish 1998)}
\begin{center}
\begin{tabular}{l r r r}
\hline
& $\theta$ & $\eta$ & $\zeta$ \\
& mas & mas & mas \\
\hline
\\
\multicolumn{4}{l}{{18 MSP with radio timing only}} \\
\\
{{DE405 $\rightarrow$ DE200}} & {-0.4 $\pm$ 0.3} & {-13 $\pm$ 0.4} &{ -13 $\pm$ 0.3} \\
{{DE405 $\rightarrow$ DE200 \cite{DE405}}} & {-1 $\pm$ 2} & {-14 $\pm$ 3} & {-10 $\pm$ 3} \\
\\
{DE414 $\rightarrow$ DE405} & {1.5 $\pm$ 0.3} & {-1.0 $\pm$ 0.4} &{ -0.9 $\pm$ 0.3} \\
{DE421 $\rightarrow$ DE405} & {1.5 $\pm$ 0.3} & {-0.9 $\pm$ 0.4} &{ -0.8 $\pm$ 0.3} \\
{INPOP08 $\rightarrow$ DE405} & {1.3 $\pm$ 0.3} & {-0.3 $\pm$ 0.4} &{ -1.1 $\pm$ 0.3} \\
{{INPOP10A $\rightarrow$ DE405}} & {1.6 $\pm$ 0.3} & {-0.7 $\pm$ 0.4} &{ -0.7 $\pm$ 0.3} \\
\\
\multicolumn{4}{l}{{J0737-30, J1713+07, B1937+21, J2145-07 with radio timing only}} \\
\\
{DE405 $\rightarrow$ DE200} & { -0.5 $\pm$ 0.2} & { -12 $\pm$ 0.3} & {-13 $\pm$ 0.18} \\
{INPOP08 $\rightarrow$ DE405} & {1.4 $\pm$ 0.03} & {-0.03 $\pm$ 0.05} & {-1.4 $\pm$ 0.03} \\
{INPOP10a $\rightarrow$ DE405} & {1.7 $\pm$ 0.01} & {-0.03 $\pm$ 0.02} & {-1.0 $\pm$ 0.01} \\
\\
\multicolumn{4}{l}{{J0737-30, J1713+07, B1937+21, J2145-07 with radio timing + VLBI}} \\
\\
DE200 $\rightarrow$ ICRF & {6 $\pm$ 4} & {26 $\pm$ 9} & {9 $\pm$ 5 }\\
DE200 $\rightarrow$ ICRF \cite{1994A&A...287..279F} & {2 $\pm$ 2} & {12 $\pm$ 3} & {6 $\pm$ 3} \\
\\
{DE405 $\rightarrow$ ICRF} & {6 $\pm$ 4} & {14 $\pm$ 9} & {-4 $\pm$ 5 }\\
{INPOP08 $\rightarrow$ ICRF} & {4 $\pm$ 4} & {14 $\pm$ 9} & {-2 $\pm$ 5 }\\
INPOP10a $\rightarrow$ ICRF & {4 $\pm$ 4} & {14 $\pm$ 9} & {-2.5 $\pm$ 5 }\\
DE421 $\rightarrow$ ICRF & {4 $\pm$ 4} & {14 $\pm$ 9} & {-3.0 $\pm$ 5 }\\
\hline
\end{tabular}
%\normalize
\label{ICRFangles}
\end{center}
\end{table}

\section{Acknowledgments}
This work was supported by the Centre National d'Etudes Spatiales (CNES) and by the Plan Pluriformation 2008-2012 of the French National Education Ministery.

%\begin{figure}
%\begin{center}
\includepdf[pages=-]{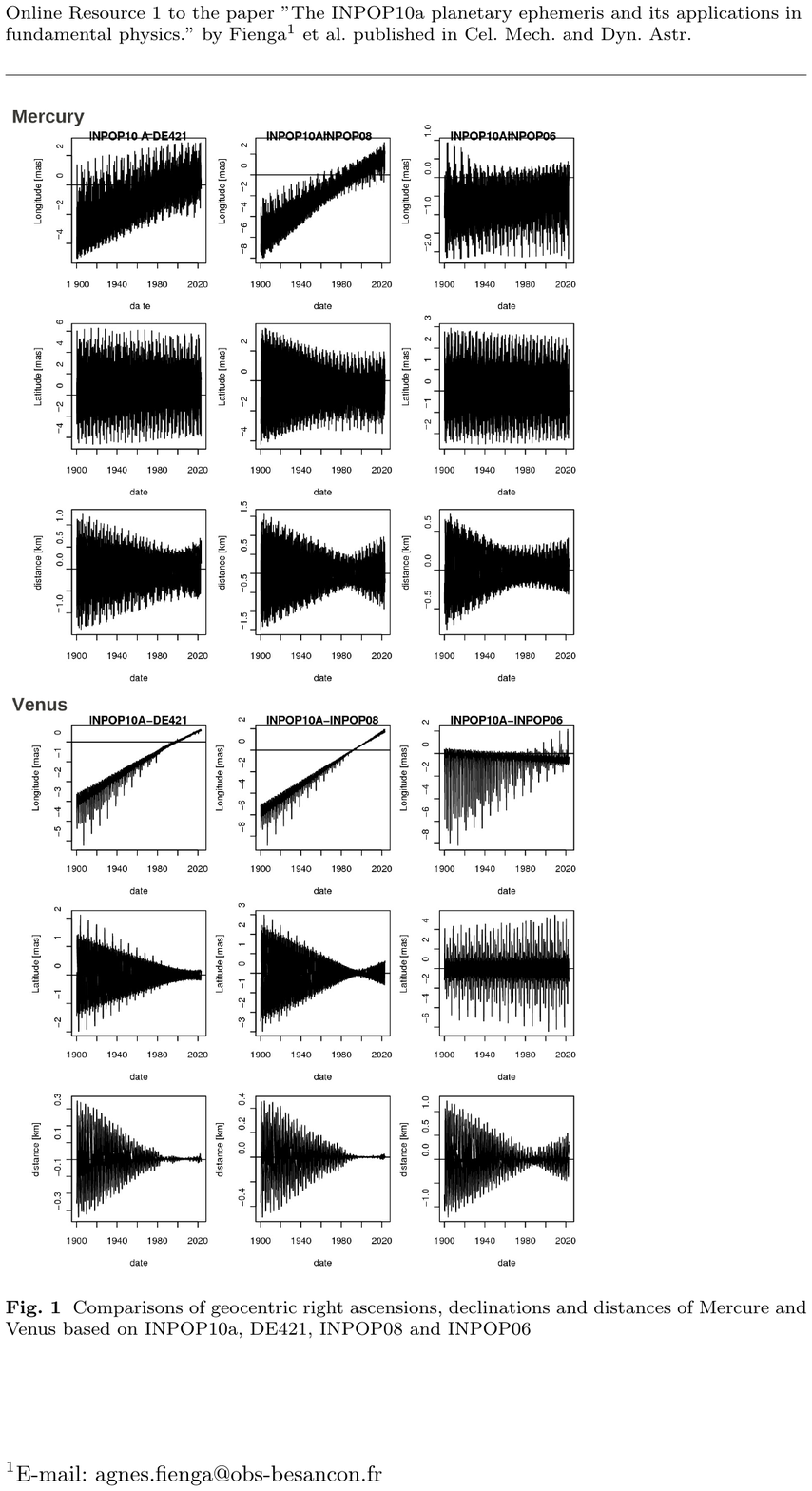}

\includepdf[pages=-]{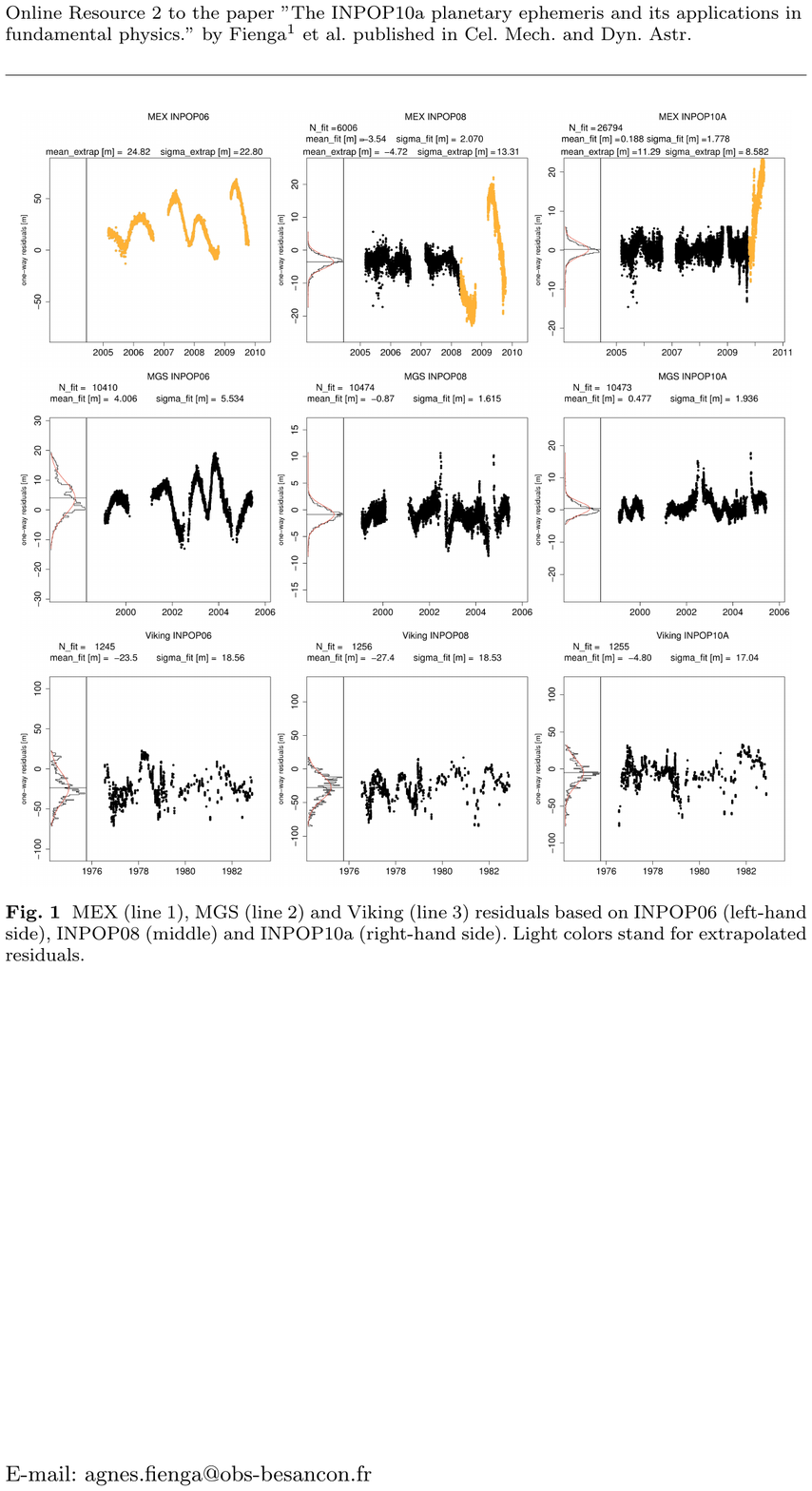}

\includepdf[pages=-]{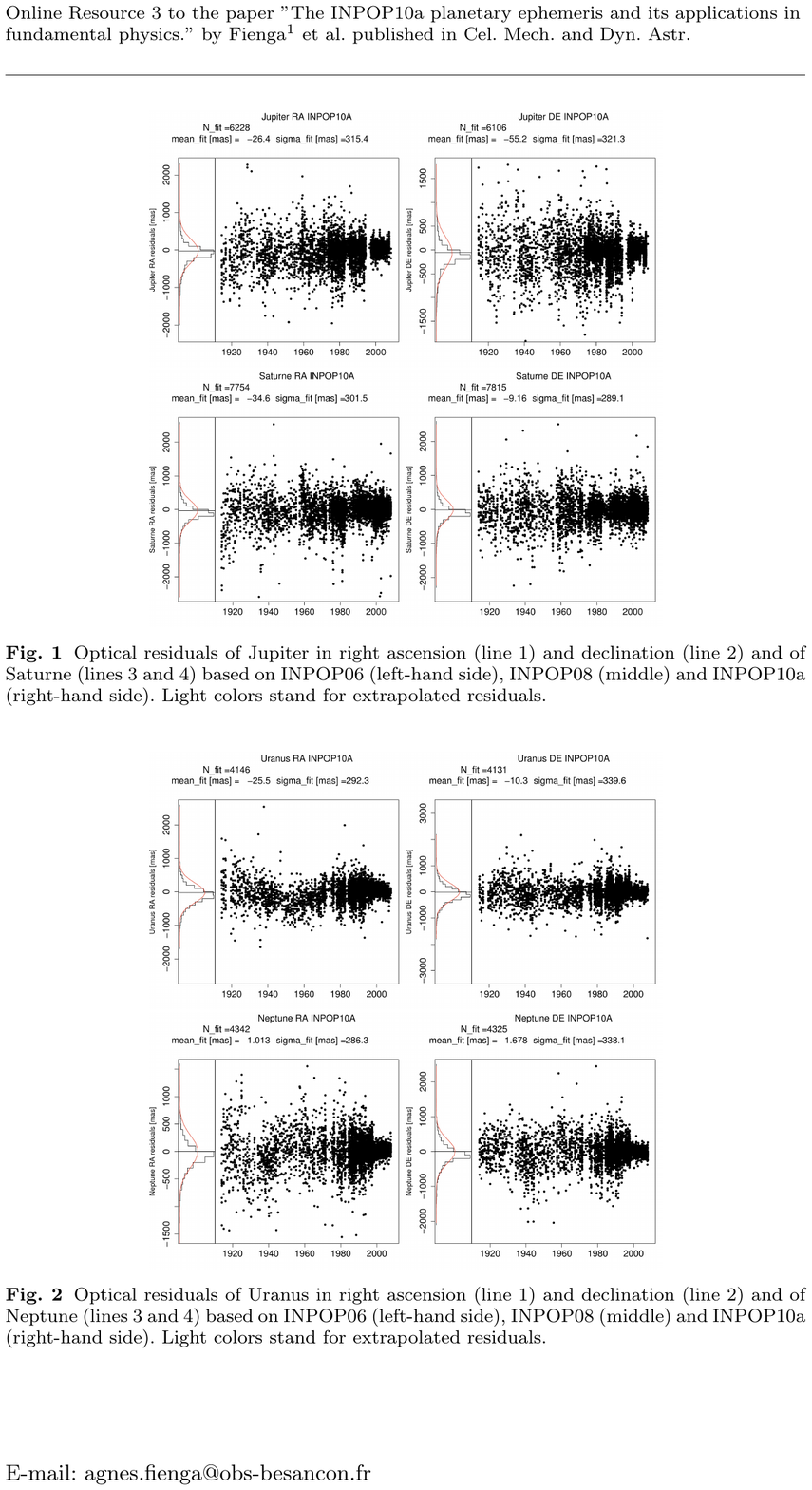}
%\end{figure}

\section{REFERENCES}
% Please type the reference as follows
% Name Initial, year, "title", journal, vol. , pp. x-x.
%
% Examples:
%
% Author1, N., Author2, N.: Title of the paper. \aap, {\bf 111}, pp. 111--222 (2000).
%
% Author2, N., Author3, N., 2003, ``Title of the paper'',
% \jgr (Solid  Earth), 111(B5), doi: 10.1000/2002JB001111.
%
% PLEASE DO NOT USE ANY SPECIAL FONTS 
% (no italics, no boldface, etc.)
%

\bibliography{biblio_these}{}
\bibliographystyle{plain}

\end{document}